\documentclass{aa} 
\usepackage{natbib}
\usepackage{graphicx}
\usepackage{txfonts}
\usepackage{epsfig}
\usepackage{times}
\usepackage{rotating}
\usepackage{float}
\usepackage[caption = false]{subfig}
\usepackage{setspace}
\usepackage{lscape}
\usepackage{epstopdf}
\usepackage{tabularx}
\usepackage{caption}
\usepackage[usenames]{color}
\usepackage[toc,page]{appendix}
\usepackage{soul}
\usepackage{amsmath}

\begin{document}

\title{Galaxy Transformation Across the Cosmic Web: Evolution of Stellar Colours and Star Formation Rates in Filaments}

\authorrunning{S. Zarattini \& J. A. L. Aguerri}

\author{S. Zarattini\inst{1,2,3} \& J. A. L. Aguerri\inst{2,3}}

\institute{Centro de Estudios de F\'isica del Cosmos de Arag\'on (CEFCA), Unidad Asociada al CSIC, Plaza San Juan 1, 44001 Teruel, Spain
 \and Instituto de Astrof\'isica de Canarias, calle Vía L\'actea s/n, E-38205 La Laguna, Tenerife, Spain
\and Departamento de Astrof\'isica, Universidad de La Laguna, Avenida Astrof\'isico Francisco S\'anchez s/n, E-38206 La Laguna, Spain\\
szarattini@cefca.es}

\date{\today}

\abstract{Galaxies form and evolve within the diverse environments of the cosmic web, which shape their properties in unique ways. However, disentangling the effects of internal properties, local environments, and large-scale structures on galaxy evolution involves significant complexities, since these effects are frequently overlapping.}
{The aim of this work is to provide evidence of the imprints left by large-scale structure on galaxy properties by selecting extensive samples of galaxies from different environments, while matching their intrinsic and local environmental characteristics. }
{We investigate the effects of the large-scale structure on the $g - r$ stellar colour and star formation rate (SFR) in galaxies with stellar mass $M_{*} > 10^{10} \ M_{\odot}$, within the redshift range $0.05 \leq z \leq 0.1$, and selected from the main galaxy sample of the Sloan Digital Sky Survey Data Release 16. We use a spectroscopic catalogue available in the literature to define samples of galaxies located in field and filament environments.}
{Galaxies located in the field tend to exhibit bluer $g - r$ stellar colours and higher SFR than those in filaments. These differences persist even in samples of galaxies matched by mass and galaxy local overdensity, indicating that they are not produced by internal or local environmental processes. These differences cannot be attributed to variations in morphology between the two matched samples.  There is also a variation in stellar colour and SFR with distance to the filaments ($D_{fila}$). The SFR of galaxies becomes statistically smaller than that of field galaxies for objects located at $D_{fila} < 5$ Mpc, while changes in stellar colour occur at smaller distances ($D_{fila} < 1$ Mpc). This could indicates that the typical filament width is about 2.5 - 5 Mpc. }
{The variations in colour and SFR between galaxy samples in the field and in filaments, with matched masses and local overdensities, indicate that large-scale environmental factors drive these transformations rather than local or internal galaxy properties. These transformations are not strong enough to produce changes in galaxy morphology. They could be explained by a cosmic web starvation process, in which galaxies moving from the field into filaments become detached from their gas supply, leading to a gradual decline in SFR and a subsequent reddening of their stellar populations. }


\authorrunning{S. Zarattini,  J. A. L. Aguerri}
\titlerunning{Galaxy Transformation Through the Cosmic Web}
\maketitle

\section{Introduction}
\label{sec:intro}

The Cold Dark Matter (CDM) paradigm states that the large-scale distribution of matter in the Universe on megaparsec scales is not homogeneous. Instead, it forms a cosmic web consisting of galaxy filaments, walls, voids, and nodes. Galaxies form and evolve within these different environments, which directly influence their properties throughout their lifetimes \citep[see e.g.][]{Kraljic2018, Xu2020}.

The primary factor shaping this cosmic web is the influence on the matter of the large-scale tidal field. Subsequent gravitational collapse amplifies anisotropies in the primordial matter distribution, resulting in a today Universe with highly asymmetric structures \citep[][]{Zeldovich1970}. The cosmic structure, predicted by theory, has also been validated by N-body numerical simulations \citep[see e.g.][]{White1987, Springel2005, Dolag2006}. Moreover, extensive spectroscopic surveys, such as the Sloan Digital Sky Survey \citep[SDSS,][]{York2000}, the 2dF Galaxy Redshift Survey \citep[2dFGRS,][]{Colless2001}, the Galaxy And Mass Assembly survey \citep[GAMA,][]{Driver2009}, and the 2MASS redshift survey \citep[2MASS,][]{Huchra2012}, have observationally confirmed the theoretical predictions regarding the structure of the cosmic web.

Galaxy evolution is strongly driven by gravity through gravitational interactions, both local and global. Local gravitational interactions occur in typical scales of smaller than 1 Mpc due to nearby companions or within the local galactic environment. These interactions can significantly impact a galaxy's morphology, dynamics, and stellar/gas content. Several types of local gravitational interactions can affect galaxies, including galaxy mergers, tidal interactions, satellite accretion, and interactions with the hot intracluster medium \citep[][]{Blanton2009,Schaye2015}

Galaxy mergers are crucial for understanding galaxy evolution. The current cosmological paradigm of galaxy formation establishes that galaxies are assembled in a bottom-up scenario, through the mergers of smaller halos \citep[][]{White1978,White1991}. Mergers between galaxies of similar mass typically result in the formation of massive spheroidal galaxies, leading to significant transformations in the morphologies of the parent galaxies \citep[see e.g.][]{Toomre1972, Somerville2015}. These interactions can also trigger intense star formation in the resulting galaxy due to the rapid consumption of gas from the pre-merger galaxies \citep[][]{Barnes1996, Kim2009}. However, other studies such as \cite{Knapen2015} or \cite{Pearson2019} did not find this star formation triggering. The merger processes have often been invoked to explain the observed dependence of galaxy morphology \citep[][]{Dressler1980}, stellar colour \citep[][]{Baldry2006, Bamford2009}, and star formation \citep[][]{Kauffmann2004} on the local environment. Conversely, mergers between galaxies of vastly different masses—often referred to as satellite accretion—slightly affect the structure of the more massive galaxy while mixing the gas and stars of the less massive galaxy into the larger one \citep[][]{Aguerri2001, Elichemoral2006, Elichemoral2011}.

When galaxies pass near one another, they experience tidal gravitational fields that can alter their morphology, resulting in long tidal tails or even inducing structures such as bars \citep[][]{Hibbard1995,Lokas2016, Martinezvalpuesta2017} or lopsided discs \citep[][]{Lokas2021, Lokas2022}. Repeated fast tidal encounters produce what is called galaxy harassment which can also shape the morphology of the galaxies \citep[][]{Moore1996, Moore1998}. Additionally, interactions among galaxies in local environments, such as galaxy clusters, can lead to variations in their cold gas content \citep[through processes like ram pressure stripping, ][]{Gunn1972, Quilis2000} or their hot gas content \citep[due to starvation or strangulation ][]{Bekki2002, Fujita2004}, ultimately reducing their star formation rates.

On scales of several megaparsecs, galaxies are embedded in the cosmic web, and its gravitational field can also influence their properties. About 35-40 $\%$ of the total galaxy luminosity is located in filaments \citep[][]{Tempel2014}. These galaxies may experience morphological and stellar content transformations due to interactions with nearby galaxies, the gas present within the filaments and/or their orientation respect to the large-scale structure \citep[][]{Dubois2014}. Additionally, the presence of filaments can affect the accretion of gas into galaxies, slowing down their star formation. These processes can lead to a reduction or suppression of star formation in galaxies within filaments \citep[][]{Aragoncalvo2019,Bulichi2024}.

Several observational studies have found the imprints left by large-scale structure on the properties of galaxies. In particular, the fraction of red galaxies, for a given stellar mass, turned out to be higher in filaments than in the field \citep[][]{Kraljic2018, Laigle2018, Pandey2020, Hoosain2024}, which contributes to galaxies displaying redder stellar colours near filaments \citep[][]{Rojas2004, Kuutma2017, Luber2019}. Moreover, \cite{Kuutma2017} found that the Elliptical-to-Spiral ratio varies with the distant to filaments, being larger at smaller distances to filaments. \cite{Chen2017} discover that galaxies closer to filaments are redder, larger and more massive than those located at large distances. Some of this galaxy properties observed in the nearby Universe are kept at higher redshift. In this sense, \citet{Malavasi2017} found that the most massive and quiescent galaxies are closer to the filaments in the VIPERS survey at $z \sim 0.7$. These observational findings have recently been confirmed analysing several galaxy properties in SIMBA simulation \citep[][]{Bulichi2024}. In addition to these morphological properties, it has been also observed that there is a correlation of the spin and principal axes of the galaxies with nearby filaments. This could be related with the acquisition of the angular momentum by the baryons in galaxies \citep[][]{Tempel2013, Zhang2013}. 

However, \citet{Kotecha2022} found an opposite trend in the central regions of galaxy clusters. In fact, these authors claimed that, within one virial radius from clusters of The Three Hundred project simulation, galaxies in filaments are bluer and more star forming than their counterparts far from filaments. This result would be associated with streams of cold gas moving together with galaxies in the filaments, that are able to prevent strangulation by shielding the effect of the hot intra-cluster medium.
\citet{Das2023} studied the effect of filaments on galaxy pairs, finding that close pairs are bluer and more star forming in filaments, whereas the trend inverts for large pairs, with separation > 50 kpc.

The large-scale structure can not only influence intrinsic galaxy properties but also produce significant effects on the efficiency of galaxy formation. Specifically, \citet{Guo2015} found that the satellite luminosity function (LF) of galaxies in filaments is higher (i.e. more satellites) than that of galaxies not located in filaments.

For years, there has been ongoing debate about whether the observed variations in galaxy properties are primarily driven by internal factors or external environmental influences. Some studies suggested that internal properties, like stellar mass and morphology, play a dominant role in shaping galaxy characteristics \citep[][]{Alpaslan2015, Wegner2019}. On the other hand, research by \citet{Eardley2015} argued that environmental factors, such as local galaxy density and position within the cosmic web, are significant drivers of these variations. Recently,  \cite{Okane2024} analyse the morphology and star formation of galaxies of the nearby Universe in different environments from voids to cluster cores. They found that galaxies in filaments tend to be less star forming and favour early-type morphologies. However, these differences observed between the population of galaxies in filaments and the field vanish when they compare samples of galaxies matching their mass and local density. This points toward the difficulty to split between different processes that simultaneously influence galaxies. Although internal and external processes act at the same time, they could have very different time scales. The results presented in  \cite{Okane2024} highlight the crucial role that sample selection plays in these research outcomes. This finding also emphasizes the importance of analysing galaxy samples with matched properties to effectively distinguish between the influences on galaxy characteristics resulting from local versus large-scale processes.

The aim of this paper is to analyse the influence of the cosmic web environment on various galaxy properties. In particular, we study the stellar colour and star formation rate (SFR) of a large number of galaxies ($\approx 120\,000$) with stellar masses greater than $10^{10} M_{\odot}$ within the redshift range of $0.05 < z < 0.1$, located in different cosmic web environments. Special attention will be given to the selection of samples with similar mass and local density properties. Matching in mass help us avoid differences arising from internal processes within the galaxies, while matching overdensities eliminate variations in galaxy properties due to different local environments. 

This work is organized as follows: the sample and the selection of the different environments are presented in Sect. 2. The impact of the large-scale environments on galaxy properties are shown in Sect. 3 . The discussion and conclusions are given in Sect. 4 and 5, respectively. 

Throughout this paper we have used the $\Lambda$CDM cosmology with $\Omega_{m} = 0.3$, $\Omega_{\Lambda} = 0.7$, and $H_{0} = 70$ km s$^{-1}$ Mpc$^{-1}$.

\begin{figure}
    \centering
    \includegraphics[width=0.5\textwidth, trim=0 0 0 0]{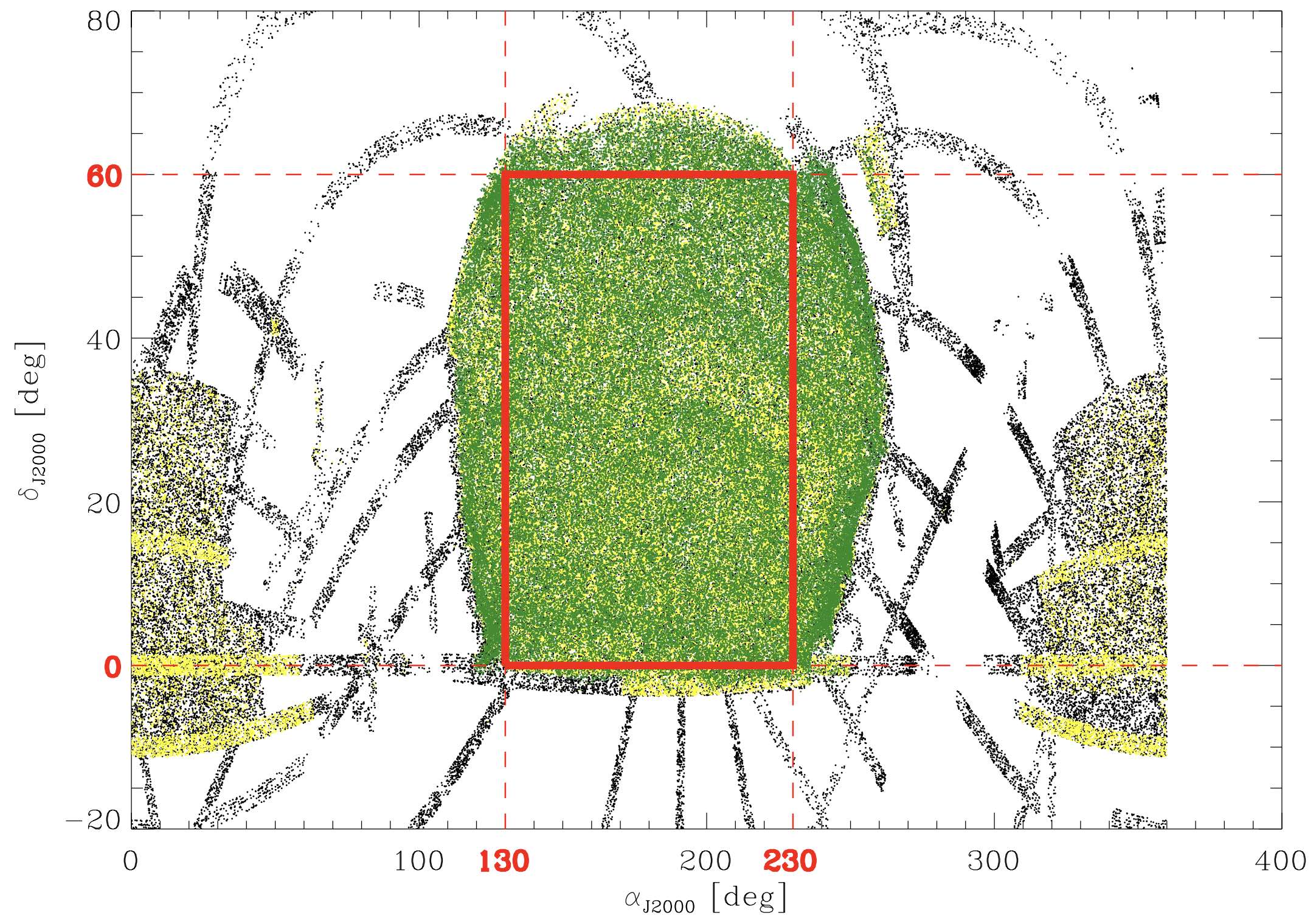}
    \caption{The SDSS footprint represented with black dots. In yellow, we show the galaxies with spectroscopic redshift. In green, we over plotted the region analysed by \citet{Chen2016}. Finally, the area delimited by the red thick solid lines is the one used in this work and the corresponding $\alpha_{J2000}$ and $\delta_{J2000}$ values are highlighted in red in the axes. Please note that we only rendered a random subsample of points for each colour. This was done for display porpoises only.}
    \label{fig:footprint}
\end{figure}

\section{Sample of galaxies and their cosmic web location}
\label{sec:sample}
In this section we present the mother galaxy sample and the cosmic web catalogue that we use to define the large-scale structure. Moreover, we discuss how we associate galaxies with different environments (filaments, field, and intersections) and how we estimate the local density of galaxies.

\subsection{Mother galaxy sample}
\label{sec:mothersample}
We use data from the SDSS Data Release 16 \citep[DR16,][]{Ahumada2020} and, in particular, we use all objects catalogued as galaxies brighter than $r_p = 17.77$, where $r_p$ is the $r-$band Petrosian magnitude. This is the same magnitude limit used to select the SDSS spectroscopic sample called main galaxy sample \citep{Strauss2002}. The separation between galaxies and stars in the Sloan Survey is based on several photometric properties of the objects. This galaxy/star classification is not free of misclassifications. For this reason, we filter the objects classified as galaxies in the SDSS database by removing those that, based on their position in the $r \ - <\mu(r_{50})>$ plane, are inconsistent with being galaxies \citep[see e.g., ][]{Sanchez-Janssen2005,Aguerri2020}. Here, $<\mu(r_{50})>$ represents the mean surface brightness of the object within its effective radius ($r_{50}$). 

The final selected galaxies span a large area in right ascension ($\alpha_{J2000}$) and declination ($\delta_{J2000}$). This area is not regular, as shown with black dots in Fig. \ref{fig:footprint}. In the same figure we highlighted in yellow galaxies with spectroscopic redshift in SDSS. Finally, green dots represent the area cover by the \citet{Chen2016} catalogue of filaments and intersections. This catalogue has been used to identify galaxies within the cosmic web, and is briefly described in Sect. \ref{sec:chen}.

Since our goal is to study the dependence of the stellar colour and star formation rate of galaxies as a function of environment (local density and distance to filament, in particular), we select our final sample in a rectangular region that is included within the red thick lines in Fig. \ref{fig:footprint}. This selection ensures us to cover a uniform area in the survey and avoid border effects in the computation of the local galaxy density and the distance to filaments.

\begin{figure}
    \centering
    \includegraphics[width=0.5\textwidth]{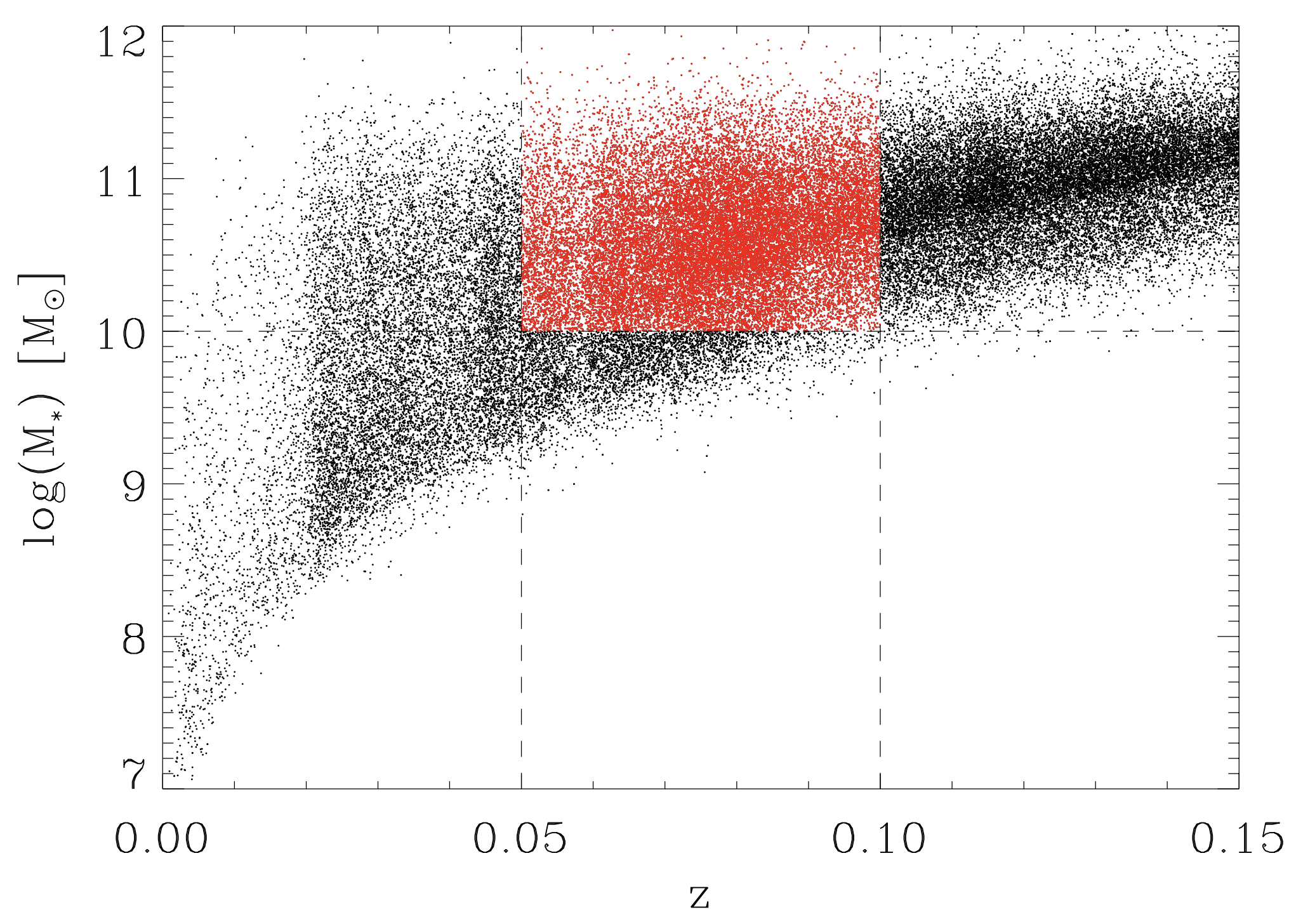}
    \caption{The dependence of the stellar mass with redshift for the SDSS sample with $r_p < 17.77$. In red, our final sample of galaxies with $0.05 \leq z \leq 0.1$ and  $M_{*} \geq 10^{10}$ M$_\odot$.}
    \label{fig:z-mass}
\end{figure}

We also applied a redshift cut, specifically excluding all galaxies with $z < 0.05$, as the filaments and intersections catalogue is defined within the range $0.05 \leq z \leq 0.7$. We then restricted our sample to $z \leq 0.1$ to ensure mass completeness.  Figure \ref{fig:z-mass} illustrates the dependence of mass on redshift for our galaxy sample, with mass estimates derived from colours following \citet{Roediger2015}. The adopted redshift cuts provide a galaxy sample that is complete in stellar mass down to $M_{*} = 10^{10}$ M$_{\odot}$. These galaxies, highlighted in red in Fig. \ref{fig:z-mass}, represent our final sample, totalling 120\,927 galaxies.  

\subsection{The cosmic web catalogue}
\label{sec:chen}
As we already did in previous works \citep{Zarattini2022,Zarattini2023} we used the catalogue of filaments and intersections presented in \citet{Chen2016} as the reference for computing distances between our galaxies and such large-scale structures.

\citet{Chen2016} constructed the catalogue using the Subspace Constrained Mean Shift (SCMS) method, which consists of multiple two-dimensional maps representing filaments and intersections, beginning at a redshift of 0.05. Each map is generated by grouping galaxies within redshift slices of $\Delta z = 0.005$ up to a redshift limit of $z=0.7$. This slicing was performed for several reasons, including the reduction of the so-called finger-of-god effect \citep[see][for details]{Chen2015}. Filaments are defined as a series of points in $\alpha_{J2000}$ and $\delta_{J2000}$, outlining the filament spine. For intersections, however, a single position in $\alpha_{J2000}$ and $\delta_{J2000}$ is provided. In this catalogue, intersections are defined as points where filaments cross. Structure formation theory suggests that galaxy clusters are often located at these filament intersections. \citet{Chen2016} show that the distance from galaxy clusters to intersections is generally smaller than that from galaxies to intersections; however, not all intersections host a galaxy cluster \citep[see][]{Chen2016}.  In Sect. \ref{sec:eRosita}, we will examine the relationship between galaxy clusters and intersections in terms of local galaxy overdensity.

The cosmic web catalogue is divided into thin slices of redshift, covering the range $0.05 \leq z \leq 0.7$, and with a thickness of $0.005$ \citep{Chen2016}. As a first step, we identify the slice corresponding to the redshift of a specific galaxy.
The filaments are described with a series of points with right ascension and declination. These points are not contiguous, but they are usually dense enough to identify the spine of each filament. Since the slices are very thin, we assume that galaxies and filaments are exactly at the same
redshift and we define the distance between them as the minimum projected mutual distance ($D_{fila}$). This methodology is the same that we adopted in \citet{Zarattini2022,Zarattini2023} and distances are measured in Mpc. We calculate the distance to intersections ($D_{int}$) using the same methodology, with the only difference that, in \citet{Chen2016} catalogue, an intersection is defined as a single point (thus, its distance correspond, by definition, to the minimum distance). In other works, $D_{fila}$ is referred to as $D_{skel}$, and $D_{int}$ is denoted as $D_{node}$ \citep[see][]{Kraljic2018}.

\subsection{Local density of galaxies}
\label{sec:local_density}

\begin{figure}
    \centering
    \includegraphics[width=0.5\textwidth]{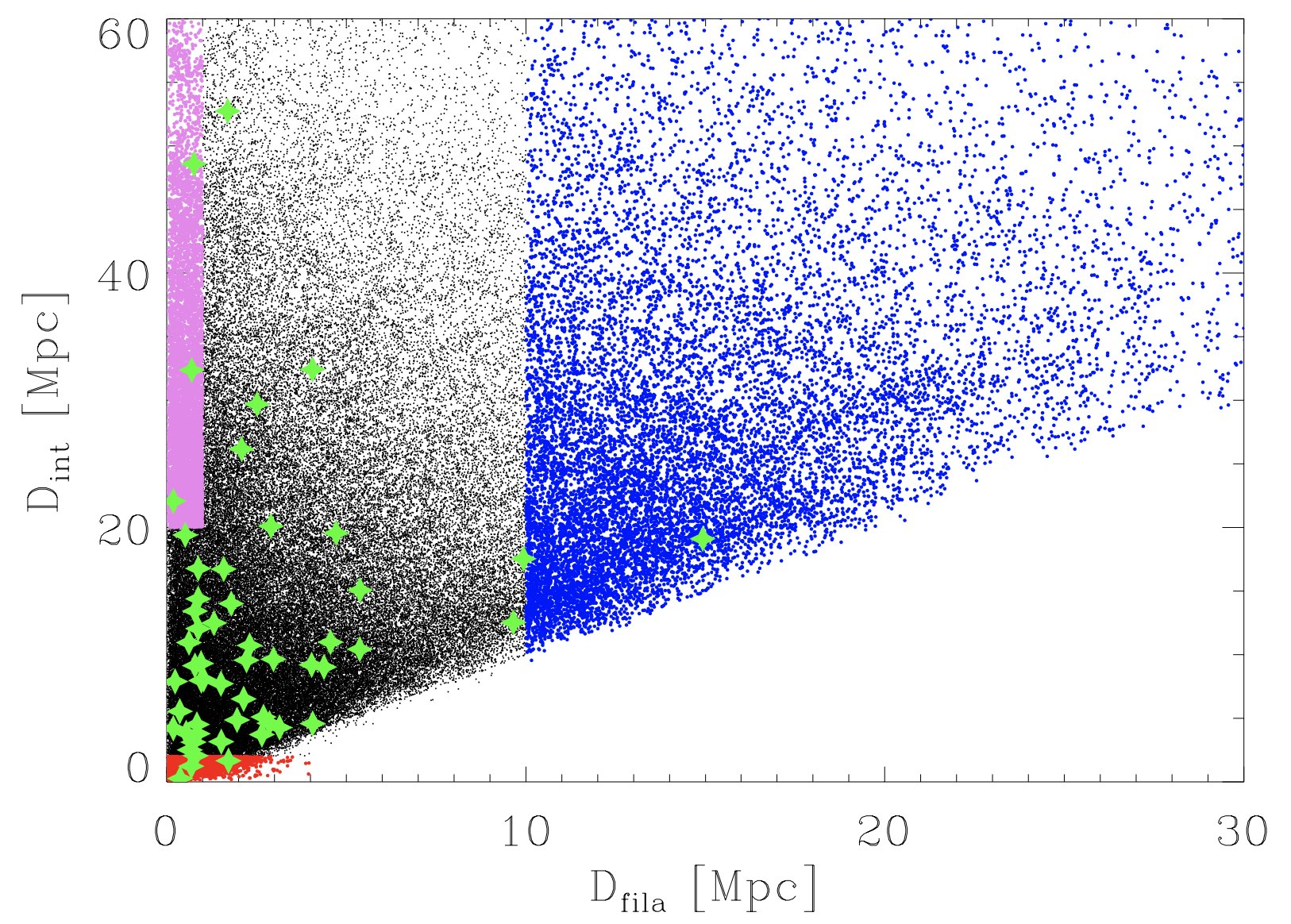}
    \caption{Distribution of the galaxies in the $D_{fila}-D_{int}$ plane. Galaxies selected in filaments, intersections and field are represented in violet, red and blue points, respectively. The green stars represent eROSITA clusters within $0.05 \leq z \leq 0.1$ and $M_{500} > 10^{14}$ M$_\odot$.}
    \label{fig:dfila_dint}
\end{figure}

One of the main drivers of galaxy evolution is the local environment, which can be traced by the projected local density of galaxies ($\Sigma_{g}$). We compute this value for each galaxy in our sample by selecting the five nearest neighbours within $\pm 3000$ km s$^{-1}$. This cut was used to properly manage the high-density environments of galaxy clusters, since it corresponds to the typical cluster escape velocity \citep{Abell1962,Pignataro2021,Zarattini2022,Boschin2023}. The spectroscopic sample of galaxies from SDSS-DR16 is incomplete, which means that the local galaxy density calculated using only spectroscopic data will be a lower limit of the real one. To account for this incompleteness and avoid underestimating the local galaxy density, we apply a photometric correction to $\Sigma_{g}$ by counting the number of galaxies without redshift in the area defined by the fifth spectroscopic neighbour. We assume that the fraction of galaxies without redshift contributing to the local density is the same as that for galaxies with spectroscopic redshift, and we add this number to the local density calculated previously.

The global mean density of galaxies in our sample  is $\overline{\Sigma}_{g} = 25.14$ galaxies per square degree, obtained as the 50\% of the cumulative distribution of local density of the entire sample ($0.05 \le z \le 0.1$). We compute the local overdensity as $\delta_{g} = (\Sigma_{g} -\overline{\Sigma}_{z,g}) / \overline{\Sigma}_{z,g}$, being $\overline{\Sigma}_{z,g}$ the mean density of galaxies in each redshift slice of $\Delta z = 0.005$, as defined in the cosmic web catalogue \citep[][]{Chen2016}. This overdensity will be used in the present work as tracer of the local galaxy environment.

\subsection{Selection of different environments}
\label{sec:environemnt}
In this section we explain how we proceed in defining the different large-scale environment used in our study.

\subsubsection{Intersections, filaments, and field}
\label{sec:int_fila_field}

We divide our sample of galaxies according to its position relative to the cosmic web. Figure \ref{fig:dfila_dint} shows the location of the galaxies in the $D_{fila}-D_{int}$ plane. We have also overploted the position in this plane of the clusters from the eROSITA  groups and cluster catalogue \citep[][]{Bulbul2024}. This catalogue is  built from 12\,247 groups and clusters detected in the X-ray band $0.3-2.3$ keV. These systems are optically confirmed and span an area of 13\,116 deg$^2$ in the Western galactic half of the sky. The redshift range is $0.003 < z < 1.32$ and the mass range is between $5\times 10^{12}$ and $2\times 10^{15}$ M$_\odot$. About two third (68\%) of the eROSITA clusters and groups are new discoveries.  For each cluster and group the catalogue provides the mass M$_{500}$, that can be converted to the M$_{200}$ mass \citep{Navarro1996} according to M$_{200} = 1.516 \, \times \, $M$_{500}$ \citep{Arnaud2005}. We can also obtain the $R_{200}$ radius by  using 

\begin{figure*}
    \centering
    \includegraphics[width=\textwidth, trim= 0 0 0 0]{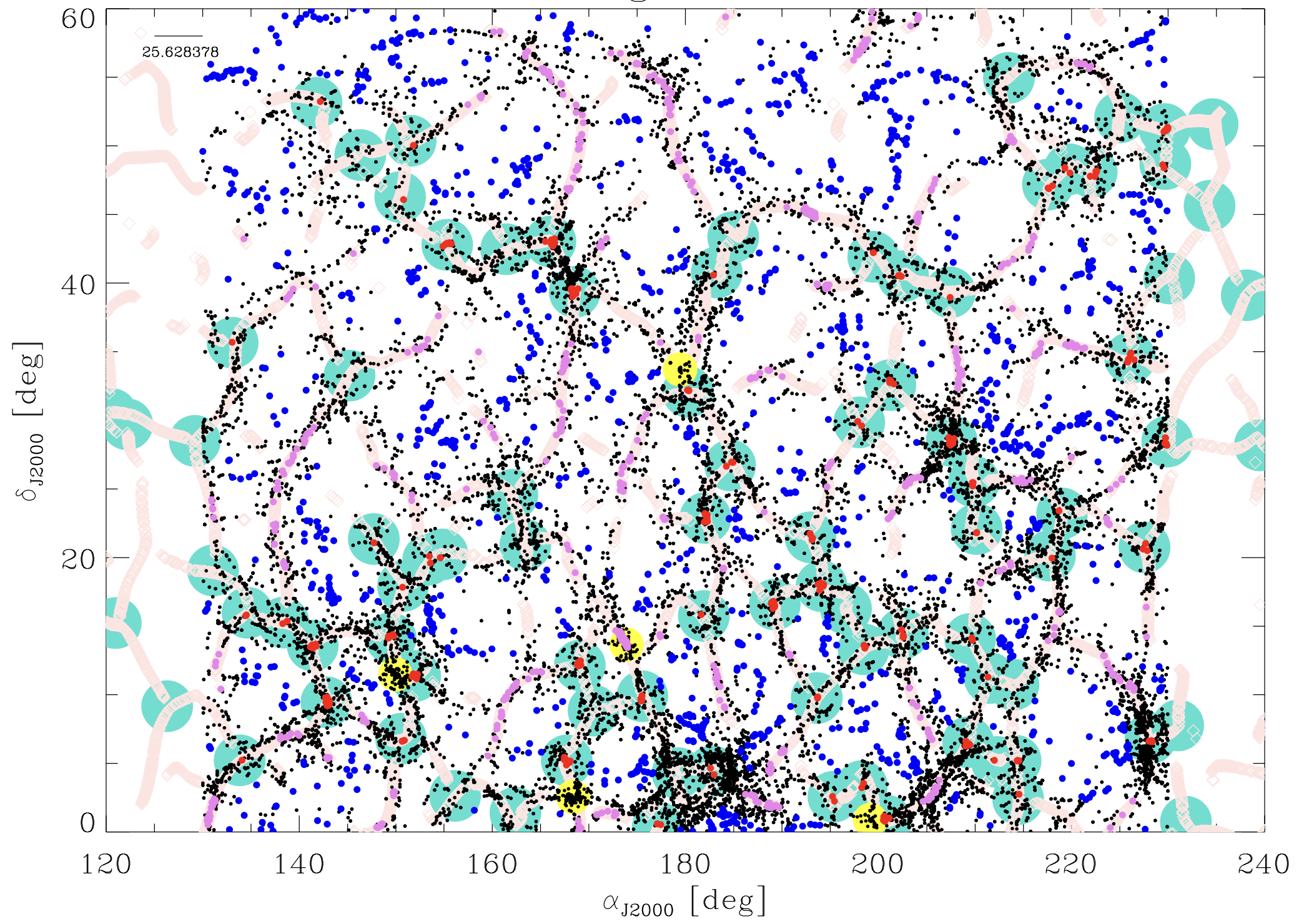}
    \caption{Plot of the entire field of view between redshift $0.075 < z < 0.08$. Red galaxies are those selected to be close to intersections ($D_{int} < 2$ Mpc), violet ones are close to filaments only ($D_{int} > 20$ Mpc and $D_{fila} < 1$ Mpc), and blue galaxies are selected to be in the field ($D_{fila} > 10$ Mpc). Turquoise circles are intersections and pink stripes are filaments from \citet{Chen2016}. In yellow we show the position of eROSITA clusters that we used in this region. Please note that symbol' sizes of filaments, intersections, and clusters are for display purposes only and have no connection with the physical sizes of these objects. The size in Mpc of the separation between two ticks (5 degrees) is indicated in the upper left corner of the image.}
    \label{fig:onefield}
\end{figure*}

\begin{equation}
{\rm M}_{200} = \frac{4}{3} \pi (200\rho_c) {\rm R}_{200}^3    
\end{equation}
where $\rho_c = 3H_0^2/8\pi G$ is the critical density of the Universe.

There are a total of 54 clusters (M$_{500} > 10^{14}$ M$_\odot$) in our footprint and with redshift between $0.05 \leq z \leq 0.1$. These are the clusters from eROSITA plotted in Fig. \ref{fig:dfila_dint}. Notice that eROSITA clusters show a wide range of $D_{int}$ values indicating that there is no a clear match between the location of these clusters and the intersections of filaments provided by \citet{Chen2016}. Following Fig. \ref{fig:dfila_dint}, we define a population of galaxies that lives close to intersections ($D_{int} < 2$ Mpc), one that reside close to filaments ($D_{int} > 20$ Mpc and $D_{fila} < 1$ Mpc), and a field population ($D_{fila} > 10$ Mpc). Taking these definitions into account, we identified $2\,963$ galaxies in intersections, $7\,042$ galaxies in filaments, and $14\,657$ galaxies in the field. These definitions were established to avoid at maximum the possible cluster contamination in the filaments and field populations (see Fig. \ref{fig:dfila_dint}). 

This division is well representative of the regions that we want to map. In Fig. \ref{fig:onefield} we show one of our redshift slices colour-coded to highlight the position of the three populations, that are clearly separated.

\begin{figure*}
    \centering
     \includegraphics[width=0.99\textwidth, trim= 0 0 0 0]{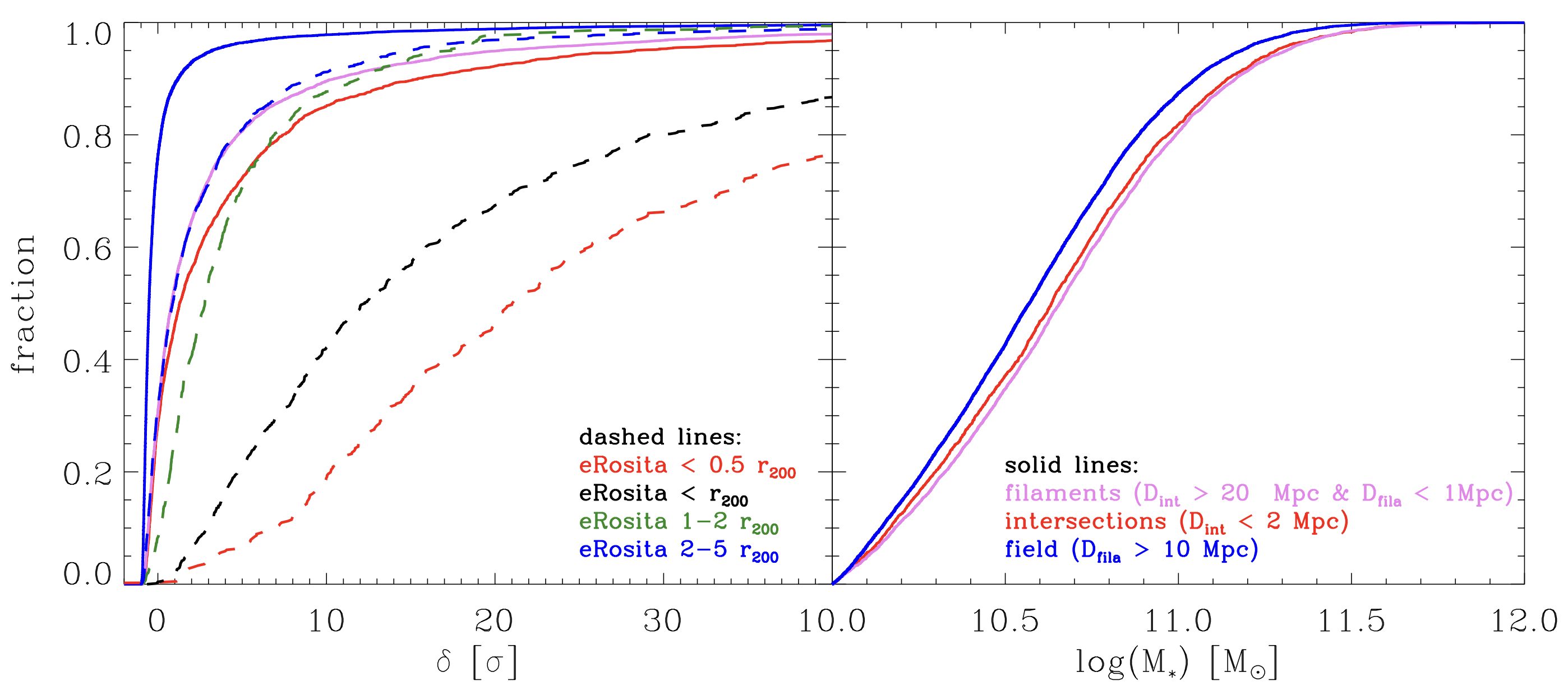}
    \caption{Left panel: the cumulative distribution of the overdensity of galaxies in the different environments. Solid lines represent the field (blue), filaments (violet), and intersections (red). The dashed lines are the eROSITA clusters and, in particular, galaxies within $2-5 \, \times$ R$_{200}$ (blue), $1-2 \, \times$ R$_{200}$ (green),$<1 \, \times$ R$_{200}$ (black), $< 0.5 \, \times$ R$_{200}$ (red). Right panel: the cumulative distribution of the stellar mass of galaxies in field (blue) filament (violet) and intersections (red).}
    \label{fig:cumulative}
\end{figure*}

The cumulative distribution of the galaxy overdensity in the different environments is shown in the left panel of Fig. \ref{fig:cumulative} using solid lines. As expected, galaxies in the field are found at the lowest overdensities with 77\% of the galaxies found in densities smaller than the mean one ($\delta_g \leq 0$), and only 5\% in $\delta_g > 3$. For the filaments, these fractions are 32\% and 28\%, respectively, whereas in intersections we find 30\% and 35\% in the same overdensity ranges. These numbers indicate a trend of increasing overdensity from the field population to the intersection regions. However, within each of the three environments, galaxies are found in distinctly different local settings. Previous works found that filaments are regions of mean overdensity where galaxy-galaxy interactions or mergers can take place \citep[][]{Guo2015,Kuutma2017, Aragoncalvo2019}. In particular, \citet{Guo2015} found that only a higher relative velocity between galaxies in the filament or a higher local density can justify the merging-rate increase observed in filaments with respect to the field. Figure \ref{fig:cumulative} also shows the galaxy overdensity for a set of galaxies located in clusters (dashed lines). These objects will be discussed in Sect. \ref{sec:eRosita}

The right panel of Fig. \ref{fig:cumulative} further illustrates the cumulative mass distribution functions of galaxies in the three environments. A clear trend in masses is visible, with smaller mass galaxies predominantly located in the field population, whereas galaxies near filaments tend to be more massive. This segregation of galaxy mass based on position within the cosmic web has also been observed by other authors \citep[][]{Chen2017,Luber2019,Hoosain2024}.

Figure \ref{fig:cumulative} clearly indicates that the galaxies located in field, filament and intersection environments  have different internal properties and are located in different local environments. This suggests that a direct comparison of galaxy properties between samples in these environments may be biased, as internal properties and local environment can become the dominant mechanisms in their evolution. 

\subsubsection{Intersections and cluster environments}
\label{sec:eRosita}

The aim of this subsection is to compare the local overdensity of galaxies at intersections with those at various distances from the centres of galaxy clusters. This comparison allows us to assess whether the galaxy population at intersections is representative of that within clusters. As we mentioned before, there are a total of 54 clusters (M$_{500} > 10^{14}$ M$_\odot$) from the eROSITA sample in our footprint and with redshift between $0.05 \leq z \leq 0.1$. We use the positions of these objects to define a galaxy population within the clusters and compare their local overdensities, up to $5 \times R_{200}$, with those of the galaxies located in the intersections of filaments. To achieve this result, we exclude clusters located within less than $5 \times R_{200}$ from the outskirts of another system. This choice reduces the number of eROSITA clusters in our footprint and redshift range to 28. The remaining objects are still sufficient, and this filtering ensures that the galaxies found in the outer regions of one cluster are not contaminated by those in the core of a nearby cluster (e.g., galaxies that would be in higher overdensities than expected in the outskirts of clusters). We add the positions of the eROSITA groups and clusters to Fig. \ref{fig:onefield} in yellow.

For this sample, we classify galaxies based on their distance from the center of the cluster (defined by the position of the eROSITA detection). The cumulative distribution of overdensities in each environment is shown in the left panel of Fig. \ref{fig:cumulative} with dashed lines. We divide the galaxies into four cluster environments: one representing the core of the clusters (distances smaller than $0.5 \, \times \, R_{200}$, 392 galaxies), galaxies within $R_{200}$ (746), galaxies between $1-2 \, \times \, R_{200}$ (659), and galaxies between $2-5 \, \times \, R_{200}$ ($1\,775$). Galaxies at the largest radii ($2-5 \, \times \, R_{200}$) are shown with a blue dashed line, and their cumulative profile is very similar to that of galaxies in filaments (as defined in the previous section). The overdensity steadily increases, but it is notable that the significant separation occurs between galaxies in the $1-2 \, \times \, R_{200}$ region and those within $R_{200}$.  
It is also important to note that galaxies located at intersections exhibit similar overdensities to those found beyond the $R_{200}$ radius of eROSITA clusters. This suggests that this population of galaxies does not represent typical virialized cluster environments but rather galaxies residing in local densities comparable to the outskirts of clusters. This could be expected, since this population was selected within a radius of 2 Mpc from the position of the intersection. In order to test this scenario, we select a galaxy population in a smaller radius, namely within 1 Mpc from the intersection (e.g. similar to the typical cluster's virial radius). The median value of the overdensity slightly grows from 1.3 to 1.4, whereas the median value of the core of eROSITA cluster is larger than 10. We thus confirm that galaxies in intersections are not representative of galaxy clusters. This is not surprising, since \citet{Chen2015} applied the SCMS algorithm to labeled simulated data generated via the Voronoi model of \citet{vandeWeygaert1994} to show that it preferentially detects structures labeled as filaments. As a consequence, we will exclude this population from further analysis and instead focus on the properties of galaxies located in field and filament environments. 

\section{Results}
In this section, we discuss our results by focusing on two main topics: the evolution of stellar $g - r$ galaxy colour and the SFR from field to filament environments.

\subsection{The impact of the environment on galaxy colours and SFR}
\label{sec:colours}

In the left panel of Fig. \ref{fig:colour-SFR} we show the cumulative distribution function of the stellar galaxy colour ($g - r$) as a function of the local environment. Solid lines are galaxies in the field (blue), and  filaments (violet). We can see that the stellar colour of galaxies in the field are bluer than in filaments.  This change can be quantified using the median values of the distributions, that is 0.74 for the field and 0.81 for the filaments. This result is similar to other previously found in the literature \citep[][]{Kuutma2017, Luber2019, Okane2024}.

To compute the SFR of the galaxies we have used the expression:

\begin{equation}
    SFR \ ({\rm M_{\odot} \, yr^{-1}}) = 7.9 \times 10^{-42} \ L_{H\alpha} \ {\rm (erg \, s^{-1}})
\end{equation}
This expression assumes a Salpeter Initial Mass Function with masses between 0.1 and 100 $M_{\odot}$ and solar metallicities \citep[see][]{Kennicutt1998}. The $H_{\alpha}$ luminosity were obtained from the continuum subtracted $H_{\alpha}$ flux  measured for their spectra by MPA-JHU \footnote{https://wwwmpa.mpa-garching.mpg.de/SDSS/DR7/} emission line analysis for the galaxies of the SDSS-DR7.  The H$_{\alpha}$ luminosity was corrected for dust absorption by using the ratio between this line and the H$_{\beta}$ line and the \cite{Cardelli1989} extinction law. In addition, the SFR obtained from the H$\alpha$ luminosity from the MPA-JHU catalog represents the fiber SFR. To obtain the total SFR  an aperture correction must be applied.  This aperture correction was done similarly to \cite{Salim2007}. The right panel of Fig. \ref{fig:colour-SFR} presents the cumulative distribution function of the SFR for galaxies located in the field and filament environments. A Kolmogorov-Smirnov (KS) test confirms that the SFR distributions of galaxies in filaments and the field are statistically distinct, with galaxies in filaments exhibiting lower SFR compared to those in the field. This difference in SFR may explain the variations observed in their stellar $g - r$ colour. The result obtained here is in agreement with \citet{Hoosain2024}. They found that galaxies nearer to filaments possess less gas, which directly results in lower SFRs. Additionally, numerical simulations show a dependence of the SFR on the distance from the filaments \citep[][]{Aragoncalvo2019, Hasan2023}. There are also some results in the literature which found evidences suggesting that interactions between galaxies within filaments can enhance their SFR \citep[][]{Darvish2014, Vulcani2019}. 

The differences observed in the stellar $g-r$ colour and SFR of galaxies in filaments versus those in the field may be influenced by the disparities in mass and local galaxy density between the two samples  (see Fig. \ref{fig:cumulative}). More massive galaxies, like those near filaments, tend to exhibit redder colours and may deplete their gas reservoirs more rapidly. Additionally, galaxies located in regions of higher local density, such as filaments, may experience gas suppression due to higher fraction of interactions, which would lead to redder colours and reduced SFR. To disentangle the effects of large-scale structures from local environmental influences, it is essential to compare galaxy samples in both field and filament environments with similar stellar masses and local galaxy densities.

\subsection{Stellar mass and overdensity-matched galaxy samples}

We construct stellar mass and local overdensity-matched samples for galaxies in filaments and the field. To achieve this, we take each galaxy from the filament environment\footnote{We select galaxies from the filament sample as it has a smaller number of galaxies than the field sample.} and find its counterpart in the field with the closest stellar mass and local overdensity. Both galaxies, from the field and filament environments, are then added to their respective matched samples. This process is repeated for all filament galaxies, ensuring no replacement of the field counterparts once selected. This procedure ensure us to have two samples of galaxies in filaments and field with similar stellar masses and overdensities. The KS test gives probabilities of 0.72 and 0.99 for the cumulative distributions of stellar mass and overdensity in the matched samples, respectively, indicating a good match between the two populations.

\begin{figure}[t]
    \centering
    \includegraphics[width=0.5\textwidth, trim= 60 400 30 230]{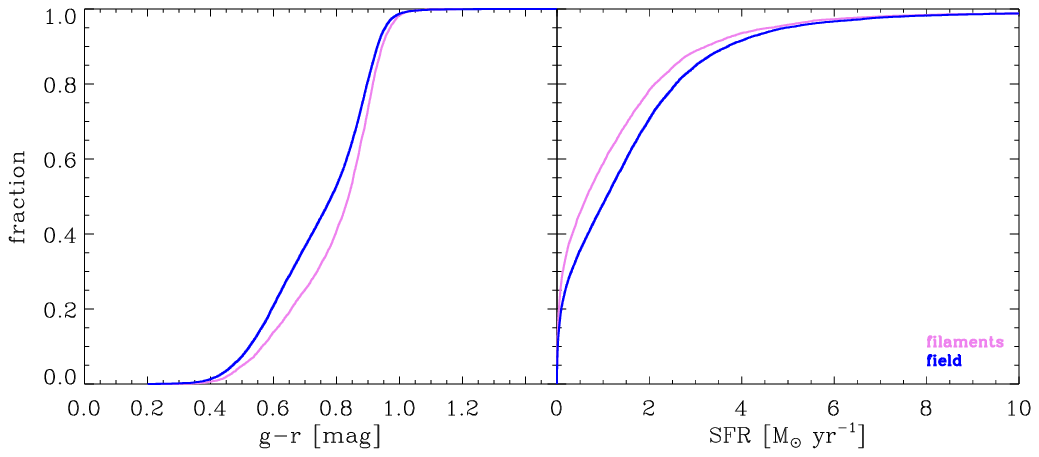}
    \caption{Left panel: cumulative distribution of the $g-r$ stellar colours of galaxies in the field (blues line), and filament (violet line). Right panel: cumulative function of the star formation rate. The colour code is the same as in left panel.}
    \label{fig:colour-SFR}
\end{figure}

\begin{figure}[t]
    \centering
    \includegraphics[width=0.5\textwidth, trim= 60 390 50 230]{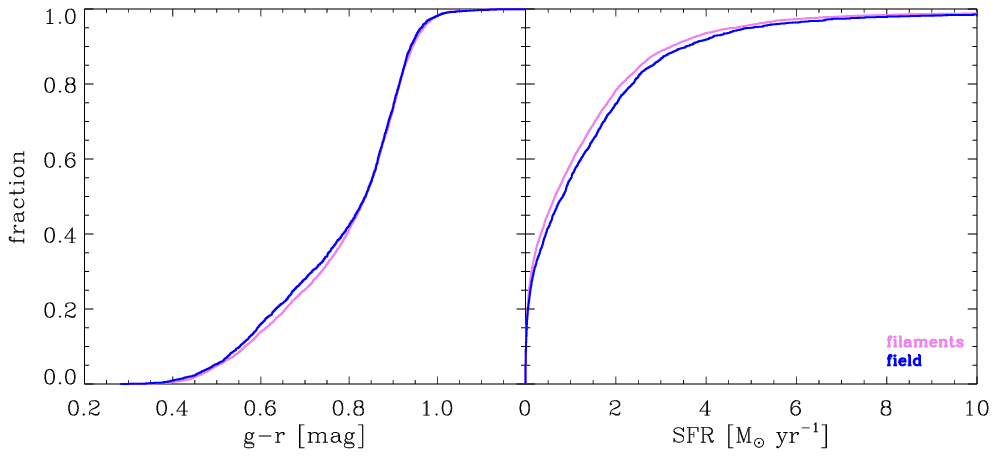}
    \caption{Cumulative distribution functions of the $g-r$ stellar colours (left panel) and SFR (right panel) of galaxies in the field (blue lines), and filaments (violet lines) for stellar mass and overdensity-matched samples.}
    \label{fig:cum_int_fila_field}
\end{figure}

The results for the stellar colour and SFR of the matched galaxy samples are presented in Fig. \ref{fig:cum_int_fila_field}. This figure demonstrates that the differences in $g - r$ stellar colour between galaxies in filaments and the field are smaller than those seen in  Fig. \ref{fig:colour-SFR}. Specifically, both distributions show similar median $g - r$ colour values. However, the KS test indicates that the two colour cumulative distributions are still statistically different ($p_{KS} = 0.002$), with the most significant difference occurring in the 50$\%$ bluest galaxies (see Fig. \ref{fig:cum_int_fila_field}). Moreover, galaxies in filaments exhibit lower SFR compared to those in the field, even in the matched samples (see Fig. \ref{fig:cum_int_fila_field}). Similar to the stellar colours, the matched samples show smaller differences in SFR than the full galaxy samples. However, the KS test gives a probability of 2.13$\times 10^{-4}$ for the cumulative SFR distribution functions of the matched samples, indicating that the two distributions are statistically different.

In summary, the analysis of stellar colour and SFR for galaxies in mass and overdensity-matched samples indicates that filaments have a measurable impact on these two properties. Specifically, galaxies in filaments are redder and exhibit lower SFR compared to their counterparts in the field.

\subsection{Change of galaxy properties with the distance to filaments}

We have found that the SFR and $g - r$ stellar colour of galaxies in filaments and in the field are statistically different, even when matching their stellar mass and local galaxy overdensity. However, we defined filament galaxies as those with $D_{fila} < 1$ Mpc and field galaxies as those with $D_{fila} > 10$ Mpc. A key question is whether there is a specific value of $D_{fila}$ at which these changes in galaxy properties occur. To investigate this, we defined three intermediate populations: $1 < D_{fila} < 2.5$ Mpc (Inter1), $2.5 < D_{fila} < 5$ Mpc (Inter2), and $5 < D_{fila} < 10$ Mpc (Inter3). All galaxies in these populations also follow $D_{int} > 20$ Mpc. These definitions give us a total of $8\,005$, $8\,395$ and $9\,589$ galaxies in the Inter1, Inter2 and Inter3 populations, respectively.

Table \ref{tab:inter_fila} presents a statistical comparison, based on the KS probability parameter $p_{KS}$, of the cumulative distribution functions of SFR and $g - r$ stellar colour between galaxies in the field and the other galaxy populations defined earlier. Notably, there is a statistically significant difference in the cumulative SFR distribution between galaxies in the field and those at $D_{fila} < 5$ Mpc. At larger radii the difference in SFR between the field and intermediate population is not statistically different at 95$\%$ c.l., but with $P_{KS}$ close to the threshold of 0.05 (see Table \ref{tab:inter_fila}). In contrast, the differences in $g - r$ colour begin for galaxies with $D_{fila} < 1$ Mpc. This suggests that, as galaxies transition from the field into filaments, they undergo a change in their star formation rate between approximately 2.5 - 5 Mpc. Meanwhile, the shift in stellar colour occurs at smaller distances from the filament ($D_{fila} < 1$ Mpc). This variation of the stellar colour with the distance to the filaments have also been observed in other galaxy samples \citep[e.g.][]{Kuutma2017}.

These findings may also indicate that the typical width of a filament could be around 2.5 - 5 Mpc, as this is the distance from the filament spine where galaxies experience changes in properties such as SFR. \citet{Castignani2022} modelled the galaxy density in the Virgo filaments using an exponential profile. Their analysis revealed a considerable diversity in the scale widths of the filaments. Notably, the widest filaments had scales ranging from 2.2 to 2.7 Mpc, likely corresponding to the types of structures identified in the \citet{Chen2016} catalogue.

In the framework where galaxies transition from the field population into the cosmic web via filaments, it seems to be a delay between the changes in their star formation rate and stellar colour. This result suggests that the SFR of galaxies changes first, likely due to a decrease in the galaxy gas content due to the lack of external gas supply, followed by a subsequent change in the stellar colour. This indicates a phased transformation where the suppression of star formation occurs before the galaxy's stellar population reddens.

\section{Discussion}

The results presented in this paper show differences in stellar colour and SFR  between samples of galaxies located in the field and in filaments, despite having similar stellar masses and local densities. This indicates that the cosmic web produce an imprint in these properties of the galaxies. 

Differences in galaxy properties between filaments and field have also been reported in the literature. For instance, several early studies have found that the fraction of red galaxies depends on their environment \citep[][]{Balogh2004,Bamford2009}. Furthermore, this fraction of red galaxies has been shown to vary with distance from filaments \citep[see ][]{Kraljic2018,Pandey2020,Hoosain2024}. Additionally, \citet{Kuutma2017} found that bright galaxies ($M_{r} \leq -20.0$) exhibit redder $g - i$ colour and lower SFR as they moves from voids to filaments. 

The differences in stellar colour and SFR between galaxies in the field and in filaments, as found in this paper, could reflect variations in the fraction of galaxy types between both samples. In this sense, \citet{Kuutma2017} observed that early-type galaxies were more abundant near filaments, which could account for the colour differences observed in their sample. To determine whether differences in morphology between field and filament environments could explain our findings, we analysed the morphology of the galaxies in the two studied environments. We obtained the morphological classification for our galaxies from \citet{Huertas-Company2011}, who used a machine-learning technique to classify the galaxies in the SDSS-DR7 into four morphological classes: Ellipticals (E), Lenticulars (S0), Early Spirals (Sab), and Late Spirals (Scd). Each galaxy in the survey is assigned a probability of belonging to each of these morphological classes. We considered a galaxy to be early-type if the combined probabilities of being elliptical (PE) and lenticular (PS0) exceeded 0.5. Otherwise, the galaxy was classified as late-type. Using this classification, 62\% of the galaxies in filaments and 64\% in the field were classified as late-type. This indicates that both samples are dominated by late-type galaxies, as expected in low-overdensity environments \citep[][]{Dressler1980}. It is also clear that the two subsamples do not present significant morphological variation. Therefore, the differences in stellar colour and SFR between galaxies in the field and in filaments cannot be explained by differences in morphology in our case. Whatever process is producing the observed differences should not be strong enough to produce significant morphological and/or structural changes in the galaxies. 

Differences in the gas mass of the galaxies in filaments and field could also explain the observed differences in stellar colour and SFR. \citet{Luber2019} found a variation of the $M_{HI}/M_{*}$ as a function of the distance to filaments. In particular, this ratio is smaller for galaxies closer to filaments. However, they also observed an increase of the stellar mass for those galaxies closer to filaments. This implies that the variation of the $M_{HI}/M_{*}$ ratio could be explained by the observed variation of the stellar mass alone.  We can compute an estimation of their gas content from the SFR of galaxies. In particular, assuming a typical depletion time $\tau = 2 \times 10^{9}$ yr for galaxies with moderate SFR and high mass as those studied here, the ratio between the mean values of the mass gas for galaxies in filament ($M_{gas,fila}$) and field ($M_{gas, field}$) is: $M_{gas,fila}/M_{gas, field} = 0.86$. Therefore, the mass of the gas in field galaxies in about 10$\%$ larger than galaxies with similar mass in filaments. This smaller amount of gas for galaxies in filaments could produce a decrement of their SFR as observed in the present work.

The decrease in SFR and gas content of galaxies from field to filaments presented in this work can be explained by several physical mechanisms. Some, such as AGN and supernova feedback processes \citep[][]{Silk1998,DiMatteo2005}, are linked to internal galaxy processes. Other mechanisms, including ram pressure stripping \citep[][]{Gunn1972,Quilis2000}, harassment \citep[][]{Moore1996}, strangulation \citep[][]{Bekki2002,Fujita2004}, or preprocessing \citep[][]{Treu2003,Goto2003}, are driven by external factors. These external mechanisms are highly dependent on both local galaxy density and mass. Therefore, in order to differentiate between the effects of the different mechanisms, it is crucial to compare samples of galaxies with matched masses and local densities as we did in the present work. \citet{Aragoncalvo2019} proposed the so-called Cosmic Web Detachment (CWD) model, which unifies several starvation mechanisms that halt star formation in galaxies due to various external processes. This model suggests that a galaxy can reduce or stop its star formation by being separated from its supply of star-forming gas. This can be achieved via galaxy-galaxy interactions, galaxy accretions to other halos or to filaments \citep[see][]{Aragoncalvo2019}. The fact that the morphology of galaxies in filaments and the field remains similar for our samples suggests that the processes affecting galaxies discard strong tidal interactions or galaxy mergers. In contrast, our results indicate that, as galaxies enter in filaments, their gas reservoirs are slowly depleted and this cosmic web depletion of the gas reservoirs in the galaxies could explain the observed differences in the stellar colour and SFR. 

In a recent study, \citet{Bulichi2024} analysed the properties of galaxies from the SIMBA simulation as a function of their distance to cosmic filaments throughout the Hubble time. They identified significant variations in stellar mass, specific star formation rate (sSFR), gas fraction, and stellar metallicity depending on how close the galaxies are to the filaments. Specifically, galaxies located near the filament spines tend to be more massive, quenched, gas-poor, and exhibit higher metallicity. 

Additionally, the depletion of gas reservoirs could also explain why the SFR of galaxies differs from that of the field sample at greater distances from the filament compared to their stellar colours. In this scenario, as galaxies approach the filaments, the infall of external gas would cease, allowing them to sustain their SFR using internal gas, but with a gradual decline. The remaining gas would be converted into stars within a closed-box framework, leading to an increase in gas metallicity due to the lack of external gas supply. Indeed, \citet{Dominguez-Gomez2023} found a slight increase of the stellar metallicity of galaxies in filaments with respect to the field ones. Moreover, significant changes in stellar colours would not occur until star formation is nearly suppressed, which would happen as galaxies move closer to the filament spine.  In this quenching process, driven by a cosmic web starvation mechanism, it is expected that the gas in field galaxies would be less metal-rich than that in galaxies within filaments. 

\begin{table}[]
    \caption{KS probabilities of the cumulative distribution functions of the field and the other galaxy populations.}
        \label{tab:inter_fila}
    \centering
    \begin{tabular}{lccccc}
    \hline
     & Filament & Inter1& Inter2 & Inter3  \\
     $D_{fila}$ [Mpc] & $<1$ & 1-2.5 & 2.5-5 & 5-10 \\
     \hline
     $P_{KS}(g-r)$ & 0.002 & 0.085 & 0.24 & 0.26\\
     $P_{KS}(SFR)$ & 4.68$\times 10^{-6}$& 0.003 & 0.006 & 0.053 \\
     \hline
     \end{tabular}
\end{table}

\citet{Peng2010} demonstrated that the effects of mass and environment on the quenching process of galaxies can be distinguished, identifying two concurrent mechanisms: mass quenching and environmental quenching. The results of this work show that there are small but statistically significant differences in their stellar colour and SFR properties when comparing samples of galaxies with similar mass and located in similar local environments (i.e., experiencing similar mass quenching and environmental quenching). The only distinction between the two galaxy samples is their cosmic web location. Therefore, the results of this study suggest that the cosmic web environment leaves a distinct imprint on the quenching process of galaxies.

\subsection{Galaxies with masses smaller than 10$^{10}$ M$_{\odot}$}
\label{sec:low_masses}

We applied a mass cut at $M = 10^{10}$ M$_{\odot}$, as Fig. \ref{fig:z-mass} indicates that the sample is mass-complete at that mass within the analysed redshift range ($0.05 < z < 0.1$). Additionally, Fig. \ref{fig:z-mass} shows that only a small number of objects with  $M < 10^{10} M_{\odot}$ can be included in our samples. Specifically, 814 and 2230 galaxies with  $M < 10^{10}$ M$_{\odot}$ are found in the filament and field samples, respectively. However, recent studies such as \citet{Galarraga2023} suggested that these galaxies can have enhanced SFR in filaments, due to the so-called cold accretion mode \citep[e.g. the gas flows directly to the centre because small galaxies are not able to support shocks, see also][]{Keres2005}.

To assess the impact of including these lower-mass galaxies, we tested our results with an extended sample incorporating all galaxies in the mass range $10^{9} M_{\odot} <$ M $< 10^{12} M_{\odot}$. We found no significant changes in the observed trends for the properties of galaxies in the filament and field samples. In both cases, galaxies in filaments exhibit redder $g-r$ colours and lower SFRs than those in the field. These differences cannot be explained by variations in morphological fractions. Furthermore, in this extended sample, galaxies in filaments also contain approximately 10$\%$ less gas than their counterparts in the field, in agreement with our main sample (M $> 10^{10} M_{\odot}$).

To better analyse the impact of filaments on galaxy properties as a function of mass, we examined the stellar colour and star formation rate (SFR) of galaxies with $10^{9} < M < 10^{10} M_{\odot}$ in both filament and field environments. These galaxies were included in the analysis done in the previous paragraph but, since they are few in number, their impact on the results could have been diluted. With this new test, we aim at highlight the properties of this population. As in our previous analysis, we matched the mass and overdensity distributions of galaxies in filaments and the field.

\begin{figure}
    \centering
    \includegraphics[width=0.5\textwidth,  trim= 170 130 180 260]{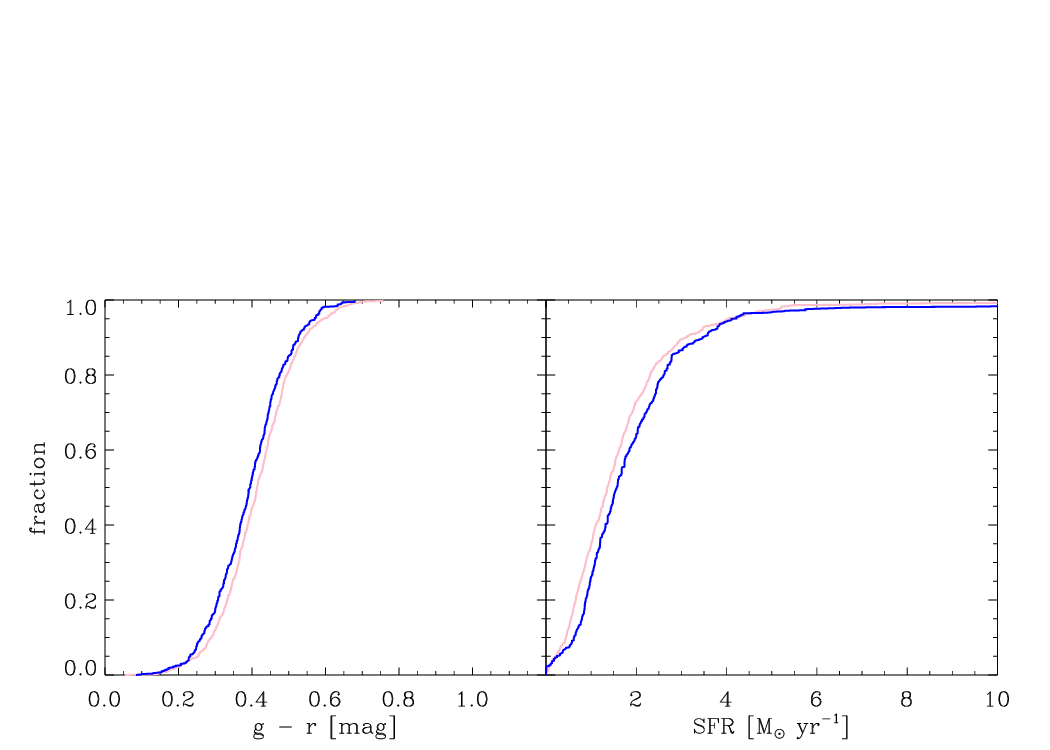}
    \caption{Left panel: cumulative distribution of the $g-r$ stellar colours of galaxies in the field (blues line), and filament (violet line). Right panel: cumulative function of the star formation rate. The colour code is the same as in left panel. In both panels, only galaxies with $10^{9} < M < 10^{10} M_{\odot}$ are considered.}
    \label{fig:colour-SFR-lowmass}
\end{figure}

Figure \ref{fig:colour-SFR-lowmass} presents the cumulative distribution functions of the $g-r$ stellar colour and SFR for the matched samples of galaxies in filaments and the field with $10^{9} < M < 10^{10} M_{\odot}$. The KS test shows that, as for our main sample of galaxies with $M>10^{10} M_{\odot}$, those located in filaments exhibit redder $g-r$ colours and lower SFRs compared to their counterparts in the field. These differences cannot be explained by variations in morphological fractions, as the fraction of late-type galaxies is 0.97 $\pm$ 0.02 in filaments and 0.99 $\pm$ 0.01 in the field.

The only relevant difference that we found is that low-mass galaxies in filaments contain approximately 20$\%$ less gas than those in the field, suggesting that they may be more affected by the large-scale structures, leading to larger gas depletion. A similar result was found in \citet{Singh2020} using the EAGLE cosmological simulation. In fact, these authors claimed that galaxies are redder and less star forming when they are closer to the spine of the filaments. These trends are stronger for galaxies less massive than $10^{10}$ M$_\odot$. In addition, \cite{Galarraga2023}, using data from the TNG50 simulation, showed that the stream connectivity of low-mass galaxies ($M \sim 10^{9.5} M_{\odot}$) depends on their large-scale environment. Galaxies in higher-density environments are connected to fewer streams compared to those in the field. This highlights the dependence of galaxy gas accretion on the surrounding environment specially for low mass galaxies.

Nonetheless, with the limited number of low-mass galaxies in our sample it is challenging to draw a definitive conclusion about the dependence of the large-scale environment on galaxy properties as a function of their stellar mass.

\subsection{Misidentified galaxies across the cosmic web}

The cosmic web is a complex network composed of various large-scale structures, including voids, walls/sheets, filaments, and clusters \citep[see e.g.,][]{Cautun2014}. Catalogs, such as the one used in this study, aim to accurately reproduce this intricate structure but are not free from uncertainties. In this work, we analyzed the properties of galaxies in two distinct environments: the field and filaments. However, these two populations may be affected by contamination from misidentified galaxies across the cosmic web.

One potential source of contamination in the filament and field galaxy populations could be the presence of galaxy clusters that were not identified as nodes or intersections in the \cite{Chen2016} catalog. As discussed in Sect. \ref{sec:eRosita}, the galaxy populations located at the intersections defined by \cite{Chen2016} do not exhibit the characteristics of typical virialized cluster environments. \cite{Chen2016} also measured the average distance between RedMapper clusters and their identified intersections. From their Fig. 11, this distance appears nearly constant with redshift, with the best agreement found in the lowest analyzed redshift range (0.10 < z < 0.15), where the difference is less than one degree (closer to 0.5, though no exact value is provided in the text). This corresponds to less than 7 Mpc at their lowest limit, z=0.1. It is worth noting that no study was done by \citet{Chen2016} in the redshift bin used in our paper. However, we were able to repeat a similar analysis using eROSITA clusters: we found a median separation of 9 Mpc between these clusters and intersections. For this reason, we think that our selection of filament galaxies (which include a distance to intersection larger than 20 Mpc) is quite
robust versus this issue. To further assess our results, we compute the total number of galaxies that are within the virial radius of an eROSITA cluster and also accomplish our definition of filament galaxy. We found that only 0.01\% of the galaxies accomplish this definition. This fraction rises to about 2\% if we extend the radius to $2 r_{200}$.
Moreover, we selected the field galaxy population as those galaxies with distance to the filaments larger than 10 Mpc. Only 1 out of 54 clusters from eROSITA is located in this region, for a fraction of galaxies smaller than 1\%. Therefore, we also believe that our selection of field galaxies is largely free from significant cluster contamination.

On the other hand, the field galaxy population may also be contaminated by galaxies residing in wall/sheet structures. These structures are challenging to detect, and a fraction of our field galaxies could indeed belong to sheets. To test this possibility, we created a subsample of field galaxies with $\delta_{g} < 0$, which should be less affected by sheet galaxies, that are expected to be found in slightly larger overdensities. This refined field sample contains 11\ 159 objects. The median and minimum distances to filaments of the galaxies in the new field sample do not change with respect to the original field ones. This implies that both galaxy samples are similarly distributed in the sky. We then compared their stellar colour and SFR properties with those of the filament population, finding no significant differences with the results presented in previous sections. Specifically, galaxies in filaments still exhibit redder $g-r$ stellar colours and lower SFRs than those in the new field sample with $\delta_{g} <0$. Additionally, the fraction of late-type galaxies remains similar between filaments and the field and the gas mass ratio between filament and field galaxies slightly changes to 0.92, indicating that filament galaxies contain about 10$\%$ less gas than those in the new field sample. These results suggest that our filament galaxy selection is robust against contamination from sheet galaxies.

\section{Conclusions}

We analysed the variation in the $g - r$ stellar colour and the SFR as a function of environment for a large sample of galaxies from SDSS-DR16 with $M_{*} \geq 10^{10}$ M$_{\odot}$ and within the redshift range  $0.05 \le z \le 0.1$. The catalogue from \citet{Chen2016} was used to classify the galaxies into three large-scale environments: field, filaments, and intersections. Additionally, we computed the local overdensity of the galaxies to account for their local environment.

There is a clear segregation of mass and overdensity among galaxies located in the field, filaments, and intersections. In particular, galaxies in filaments tend to have higher stellar masses compared to those in the field, which are generally less massive. Additionally, a gradient in local overdensity is observed, with galaxies in the field showing the lowest overdensities, followed by those in filaments, and the highest overdensities found in intersections. Galaxies in filaments inhabit environments spanning a wide range of overdensities, from those typical of field regions to others with overdensities akin to intersections. The intersections of filaments, as defined by \citet{Chen2016}, exhibit galaxy overdensities that differ significantly from those found in the virialized regions of clusters, indicating that these regions do not trace typical galaxy cluster environments.

Galaxies in the field and filaments show distinct $g - r$ stellar colours, with galaxies in filaments generally appearing redder than those in the field. Additionally, galaxies in filaments exhibit lower SFR compared to their field counterparts. These conclusions hold even for samples of galaxies in filaments and the field that are matched for stellar mass and local overdensity, indicating that filaments imprint on galaxy properties by effectively suppressing their star formation.

No differences in morphologies were found between the mass- and overdensity-matched samples of galaxies in the field and filaments. This indicates that whatever process producing the observed differences between galaxies in field and filaments is not strong enough to significantly change the morphology and/or the structure of the galaxies. In contrast, a slight variation in the gas content was observed, with galaxies in filaments containing about 10$\%$ less gas compared to those in the field. This difference rises to about 20\% for galaxies smaller than $10^{10} M_{\odot}$. These differences in gas content may contribute to the lower star formation rates seen in galaxies within filaments, as gas is a key fuel for star formation. 

There is a variation of the $g -r$ and the SFR of the galaxies as a function of the distance to the filaments. The SFR of the galaxies located at $D_{fila} < 5 $ Mpc is statistically lower than that for field galaxies. The change in the $g-r$ stellar colour is observed at closer distances to the filament ($D{fila} < 1$ Mpc).

These results indicate that filaments induce moderate changes in galaxies, such as gas suppression and the quenching of star formation, but do not lead to substantial transformations, such as major changes in galaxy morphology. This suggests that while environmental effects from the large-scale structure impact star formation and gas content, they may not be strong enough to drive significant structural changes in galaxies. Furthermore, as galaxies approach filaments, the infall of external gas would cease at distances of about 2–5 Mpc from the filament spine, gradually slowing their SFR. Significant changes in stellar colours would occur only as galaxies move closer to the filament spine ($D_{fila}<1$ Mpc) and their internal gas is depleted, ceasing the SFR and representing a galaxy quenching process driven by cosmic web starvation. In this context, differences in gas metallicity between galaxies in the field and those in filaments would also be expected. This aspect will be explored in a future study using the galaxy sample analysed in the present work.

\begin{acknowledgements}

We would like to thank to Dra. C. Muñoz-Tuñón for her valuable comments during the preparation of this work. SZ and JALA acknowledge financial support provided by the Spanish Ministerio de Ciencia, Innovación y Universidades (MICIU) through the projects PID2020-119342GB-I00 and PID2022-136598NB-C31. JALA and SZ also acknowledge support from the Agencia Estatal de Investigación del Ministerio de Ciencia, Innovación y Universidades (MCIU/AEI) under grant “WEAVE: EXPLORING THE COSMIC ORIGINAL SYMPHONY, FROM STARS TO GALAXY CLUSTERS” and the European Regional Development Fund (ERDF) with reference PID2023-153342NB-I00 / 10.13039/501100011033.
\\
SZ acknowledges the financial support provided by the Governments of Spain and Arag\'on through their general budgets and the Fondo de Inversiones de Teruel, the Aragonese Government through the Research Group E16\_23R, the Spanish Ministry of Science and Innovation and the European Union - NextGenerationEU through the Recovery and Resilience Facility project ICTS-MRR-2021-03-CEFCA, and the Spanish Ministry of Science and Innovation (MCIN/AEI/10.13039/501100011033 y FEDER, Una manera de hacer Europa) with grant PID2021-124918NB-C44.
\\
Funding for the Sloan Digital Sky Survey IV has been provided by the Alfred P. Sloan Foundation, the U.S. Department of Energy Office of Science, and the Participating Institutions. SDSS acknowledges support and resources from the Center for High-Performance Computing at the University of Utah. The SDSS web site is www.sdss4.org. SDSS is managed by the Astrophysical Research Consortium for the Participating Institutions of the SDSS Collaboration including the Brazilian Participation Group, the Carnegie Institution for Science, Carnegie Mellon University, Center for Astrophysics Harvard \& Smithsonian (CfA), the Chilean Participation Group, the French Participation Group, Instituto de Astrofísica de Canarias, The Johns Hopkins University, Kavli Institute for the Physics and Mathematics of the Universe (IPMU) / University of Tokyo, the Korean Participation Group, Lawrence Berkeley National Laboratory, Leibniz Institut für Astrophysik Potsdam (AIP), Max-Planck-Institut für Astronomie (MPIA Heidelberg), Max-Planck-Institut für Astrophysik (MPA Garching), Max-Planck-Institut für Extraterrestrische Physik (MPE), National Astronomical Observatories of China, New Mexico State University, New York University, University of Notre Dame, Observatório Nacional / MCTI, The Ohio State University, Pennsylvania State University, Shanghai Astronomical Observatory, United Kingdom Participation Group, Universidad Nacional Autónoma de México, University of Arizona, University of Colorado Boulder, University of Oxford, University of Portsmouth, University of Utah, University of Virginia, University of Washington, University of Wisconsin, Vanderbilt University, and Yale University. 
\end{acknowledgements}

\bibliography{bibliografia}
\end{document}